\documentclass{article}

\usepackage{amssymb,amsmath,amsthm,url}
\usepackage[letterpaper,hmargin=1.0in,vmargin=1.0in]{geometry}
\usepackage[pdftex,colorlinks,linkcolor=blue,citecolor=blue,filecolor=blue,urlcolor=blue]{hyperref}
\usepackage{hyperref}

\usepackage{cleveref}
\usepackage{color}
\usepackage[boxed]{algorithm}
\usepackage[margin=20pt,font=small,labelfont=bf]{caption}
\usepackage{tikz}
\usepackage{graphicx}
\usepackage{nicefrac}
\usepackage{aliascnt}
\usepackage{booktabs}

\usepackage{caption}
\usepackage{subcaption}

\newtheorem{theorem}{Theorem}[section]
\crefname{theorem}{Theorem}{Theorems}

\newaliascnt{lemma}{theorem}
\newtheorem{lemma}[lemma]{Lemma}
\aliascntresetthe{lemma}
\crefname{lemma}{Lemma}{Lemmas}

\newaliascnt{proposition}{theorem}

\aliascntresetthe{proposition}
\crefname{proposition}{Proposition}{Propositions}

\newaliascnt{corollary}{theorem}

\aliascntresetthe{corollary}
\crefname{corollary}{Corollary}{Corollaries}

\newaliascnt{fact}{theorem}

\aliascntresetthe{fact}
\crefname{fact}{Fact}{Facts}

\newaliascnt{definition}{theorem}

\aliascntresetthe{definition}
\crefname{definition}{Definition}{Definitions}

\newaliascnt{remark}{theorem}

\aliascntresetthe{remark}
\crefname{remark}{Remark}{Remarks}

\newaliascnt{conjecture}{theorem}

\aliascntresetthe{conjecture}
\crefname{conjecture}{Conjecture}{Conjectures}

\newaliascnt{claim}{theorem}

\aliascntresetthe{claim}
\crefname{claim}{Claim}{Claims}

\newaliascnt{question}{theorem}

\aliascntresetthe{question}
\crefname{question}{Question}{Questions}

\newaliascnt{exercise}{theorem}

\aliascntresetthe{exercise}
\crefname{exercise}{Exercise}{Exercises}

\newaliascnt{example}{theorem}

\aliascntresetthe{example}
\crefname{example}{Example}{Examples}

\newaliascnt{notation}{theorem}

\aliascntresetthe{notation}
\crefname{notation}{Notation}{Notations}

\newaliascnt{problem}{theorem}

\aliascntresetthe{problem}
\crefname{problem}{Problem}{Problems}

\newcommand{\norm}[1]{\lVert#1\rVert}

\def\E{\mathbb E}

\newcommand{\R}{\mathbb R}

\newcommand{\Rbb}{\mathbb{R}}
\newcommand{\eps}{\varepsilon}

\allowdisplaybreaks

\title{Scalable Nearest Neighbor Search for Optimal Transport\footnote{Code available at~\url{https://github.com/ilyaraz/ot_estimators}.}}

\author{
	Arturs Backurs\footnote{Author names are ordered alphabetically.} \\ 
	TTIC \\
	\and
  Yihe Dong\\
  Microsoft \\
  \and
  Piotr Indyk \\
  MIT \\
  \and
  Ilya Razenshteyn \\
  Microsoft Research \\
  \and
  Tal Wagner \\
  MIT \\
}

\begin{document}

\maketitle

\begin{abstract}
The Optimal Transport (a.k.a.\ Wasserstein) distance is an increasingly popular similarity measure for rich data domains, such as images or text documents.
This raises the necessity for fast nearest neighbor search algorithms according to this distance, which poses a substantial computational bottleneck on massive datasets.

In this work we introduce Flowtree, a fast and accurate approximation algorithm for the Wasserstein-$1$ distance. We formally analyze its approximation factor and running time. 
We perform extensive experimental evaluation of nearest neighbor search algorithms in the $W_1$ distance on real-world dataset. 
Our results show that compared to previous state of the art, Flowtree achieves up to $7.4$ times faster running time.
\end{abstract}

\section{Introduction}

Given a finite metric space $\mathcal{M} = (X, d_X)$ and two distributions
$\mu$ and $\nu$ on $X$, the Wasserstein-$1$ distance
(a.k.a.\ Earth Mover's Distance or Optimal Transport) between $\mu$ and $\nu$ is defined as
\begin{equation}
\label{w1_def}
W_1(\mu, \nu) = \min_{\tau} \sum_{x_1, x_2 \in X} \tau(x_1, x_2) \cdot d_X(x_1, x_2),
\end{equation}
where the minimum is taken over all distributions $\tau$ on $X \times X$
whose marginals are equal to $\mu$ and~$\nu$.\footnote{For mathematical foundations of Wasserstein distances, see~\cite{villani2003topics}.}
The Wasserstein-$1$ distance
and its variants are heavily used in applications to measure similarity in structured data domains, such as images~\cite{rubner2000earth} and natural language text~\cite{kusner2015word}.
In particular, \cite{kusner2015word} proposed the~\emph{Word Mover Distance (WMD)} for text documents. Each document is seen as a uniform distribution over the words it contains, and the underlying metric between words is given by high-dimensional word embeddings such as word2vec~\cite{mikolov2013distributed} or GloVe~\cite{pennington2014glove}.
It is shown in~\cite{kusner2015word} (see also~\cite{le2019tree,yurochkin2019hierarchical,wu2018word}) that the Wasserstein-$1$ distance between the two distributions is a high-quality measure of similarity between the associated documents.

To leverage the Wasserstein-$1$ distance for classification tasks, the above line of work uses the $k$-nearest neighbor classifier. This poses a notorious bottleneck for large datasets, necessitating the use of fast approximate similarity search algorithms.
While such algorithms are widely studied for $\ell_p$ distances (chiefly $\ell_2$; see~\cite{andoni2018approximate} for a survey), much less is known for Wasserstein distances, and a comprehensive study appears to be lacking.
In particular, two properties of the $W_1$ distance
make the nearest neighbor search problem very challenging. First,
the $W_1$ distance is fairly difficult to compute (the most common approaches are
combinatorial flow algorithms~\cite{kuhn1955hungarian} or approximate iterative methods~\cite{cuturi2013sinkhorn}). Second, the $W_1$ distance is strongly
incompatible with Euclidean (and more generally, with $\ell_p$)
geometries~\cite{bourgain1986metrical,khot2006nonembeddability,naor2007planar,andoni2008earth,
andoni2015snowflake,andoni2018sketching}, which renders many of the existing
techniques for nearest neighbor search inadequate (e.g., random projections).

In this work, we systematically study the $k$-nearest neighbor search ($k$-NNS) problem with respect to the $W_1$ distance.
In accordance with the above applications, we focus on the case where the ground set $X$ is a finite subset of $\Rbb^d$, endowed with the Euclidean distance, where $d$ can be a high dimension, and each distribution over $X$ has finite support of size at most $s$.\footnote{In the application to~\cite{kusner2015word}, $X$ is the set word embeddings of (say) all terms in the English language, and $s$ is the maximum number of terms per text document.}
Given a dataset of $n$ distributions
$\mu_1, \mu_2, \ldots, \mu_n$, 
the goal is to preprocess it, such that given a query distribution $\nu$ (also supported on $X$),
we can quickly find the $k$ distributions $\mu_i$ closest to $\nu$ in the $W_1$ distance.
To speed up search, the algorithms we consider rely on efficient estimates of the distances $W_1(\mu_i,\nu)$. This may lead to retrieving approximate nearest neighbors rather than the exact ones, which is often sufficient for practical applications.

\subsection{Prior work}\label{sec:related}
Kusner et al.~\cite{kusner2015word} sped up $k$-NNS for WMD by designing two approximations of $W_1$.
The first algorithm estimates $W_1(\mu,\nu)$ as the Euclidean distance between their respective means.
The second algorithm, called ``Relaxed WMD'' (abbrev.~R-WMD), assigns every point in the support of $\mu$ to its closest point in the support of $\nu$, and vice versa, and returns the maximum of the two assignments.
Both of these methods produce an estimate no larger than the true distance $W_1(\mu,\nu)$.
The former is much faster to compute, while the latter has a much better empirical quality of approximation.
The overall $k$-NNS pipeline in~\cite{kusner2015word} consists of the combination
of both algorithms, together with exact $W_1$ distance computation.
Recently, \cite{atasu2019linear} proposed modifications to R-WMD by instating additional capacity constraints, resulting in more accurate estimates that can be computed almost as efficiently as R-WMD.

Indyk and Thaper~\cite{indyk2003fast} studied the approximate NNS problem for the $W_1$ distance in the context of image retrieval.
Their approach capitalizes on a long line of work of~\emph{tree-based} methods, in which the given metric space is embedded at random into a tree metric.
This is a famously fruitful approach 
for many algorithmic and structural statements~\cite{bartal1996probabilistic,bartal1998approximating,charikar1998approximating,indyk2001algorithmic,gupta2003bounded,fakcharoenphol2004tight,calinescu2005approximation,mendel2006ramsey}.
It is useful in particular for Wasserstein distances, 
since the optimal flow ($\tau$ in~(\ref{w1_def})) on a tree can be computed in linear time,
and since a tree embedding of the underlying metric yields an $\ell_1$-embedding of the Wasserstein distance, as shown by ~\cite{kleinberg2002approximation,charikar2002similarity}.
This allowed~\cite{indyk2003fast} to design an efficient NNS algorithm for $W_1$ based on classical locality-sensitive hashing (LSH).
Recently, \cite{le2019tree} introduced a kernel similarity measure based on the same approach, and showed promising empirical results for additional application domains.

\subsection{Our results}\label{sec:ourresults}

\paragraph{Flowtree.}
The tree-based method used in~\cite{indyk2003fast,le2019tree} is a classic algorithm called~\emph{Quadtree}, described in detail~\Cref{sec:quadtree}. In this method, the ground metric $X$ is embedded into a random tree of hypercubes, and the cost of the optimal flow is computed with respect to the tree metric. 
We suggest a modification to this algorithm, which we call~\emph{Flowtree}: It computes the optimal~\emph{flow} on the same random tree, but evaluates the~\emph{cost} of that flow in the original ground metric. 

While this may initially seem like a small modification, it in fact leads to an algorithm with vastly different properties. 
On one hand, while both algorithms run asymptotically in time $O(s)$, Quadtree is much faster in practice. The reason is that the~\emph{cost} of the optimal flow on the tree can be computed very efficiently, without actually computing the flow itself. On the other hand, Flowtree is~\emph{dramatically more accurate}. Formally, we prove it has an asymptotically better approximation factor than Quadtree. Empirically, our experiments show that Flowtree is as accurate as state-of-the-art $O(s^2)$ time methods, while being much faster.

\paragraph{Theoretical results.}
A key difference between Flowtree and Quadtree is that the approximation quality of Flowtree is~\emph{independent of the dataset size}, i.e., of the number $n$ of distributions $\mu_1,\ldots,\mu_n$ that need to be searched. Quadtree, on the other hand, degrades in quality as $n$ grows. 
We expose this phenomenon in two senses:
\begin{itemize}
  \item\textit{Worst-case analysis:}
We prove that Flowtree reports an $O(\log^2s)$-approximate nearest neighbor w.h.p~if the input distributions are uniform, and an $O(\log (d \Phi) \cdot \log s)$-approximate nearest neighbor (where $d$ is the dimension and $\Phi$ is the coordinate range of $X$) even if they are non-uniform. 
Quadtree, on the other hand, reports an $O(\log (d \Phi) \cdot \log(sn))$-approximate nearest neighbor, and we show the dependence on $n$ is~\emph{necessary}.

  \item\textit{Random model:} We analyze a popular random data model, in which both Flowtree and Quadtree recover the exact nearest neighbor with high probability. Nonetheless, here too, we show that Flowtree's success probability is independent of $n$, while Quadtree's degrades as $n$ grows.
\end{itemize}

\paragraph{Empirical results.}
We evaluate Flowtree, as well as several baselines and state-of-the-art methods, for nearest neighbor search in the $W_1$ distance on real-world datasets.

Our first set of experiments evaluates each algorithm individually. Our results yield a sharp divide among existing algorithms: The linear time ones are very fast in practice but only moderately accurate, while the quadratic time ones are much slower but far more accurate. Flowtree forms an intermediate category: it is slower and more accurate than the other linear time algorithms, and is at least $5.5$ (and up to $30$) times faster than the quadratic time algorithms, while attaining similar or better accuracy.

The above results motivate a sequential combination of algorithms, that starts with a fast and coarse algorithm to focus on the most promising candidates nearest neighbors, and gradually refines the candidate list by slower and more accurate algorithms. Such pipelines are commonly used in practice, and in particular were used in~\cite{kusner2015word} (termed ``prefetch and prune''). 
Our second set of experiments evaluates pipelines of various algorithms. We show that incorporating Flowtree into pipelines substantially improves the overall running times, by a factor of up to $7.4$.

\section{Preliminaries: Quadtree}\label{sec:quadtree}
In this section we describe the classic Quadtree algorithm.
Its name comes from its original use in two dimensions (cf.~\cite{samet1984quadtree}), but it extends to---and has been successfully used in---various high-dimensional settings (e.g.~\cite{indyk2001algorithmic,indyk2017practical,backurs2019scalable}).
It enjoys a combination of appealing theoretical properties and amenability to fast implementation.
As it forms the basis for Flowtree, we now describe it in detail.

\paragraph{Generic Quadtree.}
Let $X\subset\Rbb^d$ be a finite set of points.
Our goal is to embed $X$ into a random tree metric, so as to approximately preserve each pairwise distance in $X$.
To simplify the description, suppose that the minimum pairwise distance in $X$ is exactly $1$, and that all points in $X$ have coordinates in $[0,\Phi]$.\footnote{This is without loss of generality, as we can set the minimum distance to $1$ by scaling, and we can shift all the points to have non-negative coordinates without changing internal distances.}

The first step is to obtain a randomly shifted hypercube that encloses all points in $X$.
To this end, let $H_0=[-\Phi,\Phi]^d$ be the hypercube with side length $2\Phi$ centered at the origin.
Let $\sigma \in \Rbb^d$ be a random vector with i.i.d.\ coordinates uniformly distributed in $[0,\Phi]$.
We shift $H_0$ by $\sigma$, obtaining the hypercube $H=[-\Phi,\Phi]^d+\sigma$. Observe that $H$ has side length $2\Phi$ and encloses $X$.
The random shift is needed in order to obtain formal guarantees for arbitrary $X$.

Now, we construct a tree of hypercubes by letting $H$ be the root, halving $H$ along each dimension, and recursing on the resulting sub-hypercubes. We add to the tree only those hypercubes that are non-empty (i.e., contain at least one point from $X$). Furthermore, we do not partition hypercubes that contain exactly one point from $X$; they become leaves. The resulting tree has at most $O(\log(d\Phi))$ levels and exactly $|X|$ leaves, one per point in $X$.\footnote{This is since the diameter of the root hypercube $H$ is $\sqrt{d}\Phi$, and the diameter of a leaf is no less than $1/2$, since by scaling the minimal distance in $X$ to $1$ we have assured that a hypercube of diameter $1/2$ contains a single point and thus becomes a leaf. Since the diameter is halved in each level, there are at most $O(\log(d\Phi))$ levels.} We number the root level as $\log\Phi+1$, and the rest of the levels are numbered downward accordingly ($\log\Phi, \log\Phi-1,\ldots$).
We set the weight of each tree edge between level $\ell+1$ and level $\ell$ to be $2^\ell$.

The resulting quadtree has $O(|X|d \cdot \log(d\Phi))$ nodes, and it is straightforward to build it in time $\widetilde{O}(|X|d \cdot \log(d\Phi))$.\footnote{Note that although the construction partitions each hypercube into $2^d$ sub-hypercubes, eliminating empty hypercubes ensures that the tree size does not depend exponentially on $d$.}

\paragraph{Wasserstein-$1$ on Quadtree.}
The tree distance between each pair $x,x'\in X$ is defined as the total edge weight on the unique path between their corresponding leaves in the quadtree.
Given two distributions $\mu,\nu$ on $X$, the Wasserstein-$1$ distance with this underlying metric (as a proxy for the Euclidean metric on $X$) admits the closed-form $\sum_v2^{\ell(v)}|\mu(v)-\nu(v)|$, where $v$ ranges over all nodes in the tree, $\ell(v)$ is the level of $v$, $\mu(v)$ is the total $\mu$-mass of points enclosed in the hypercube associated with $v$, and $\nu(v)$ is defined similarly for the $\nu$-mass.
If $\mu,\nu$ have supports of size at most $s$, then this quantity can be computed in time $O(s\cdot\log(d\Phi))$.

The above closed-form implies, in particular, that $W_1$ on the quadtree metric embeds isometrically into $\ell_1$, as originally observed by~\cite{charikar2002similarity} following~\cite{kleinberg2002approximation}.
Namely, the $\ell_1$ space has a coordinate associated with each tree node $v$, and a distribution $\mu$ is embedded in that space by setting the value of each coordinate $v$ to $2^{\ell(v)}\mu(v)$, where $\mu(v)$ is defined as above.
Furthermore, observe that if $\mu$ has support size at most $s$, then its corresponding $\ell_1$ embedding w.r.t~the tree metric has at most $sh$ non-zero entries, where $h$ is the height of the tree.
Thus, computing $W_1$ on the tree metric amounts to computing the $\ell_1$ distance between sparse vectors, which further facilitates fast implementation in practice.

\section{Flowtree}\label{sec:flowtree}
The Flowtree algorithm for $k$-NNS w.r.t.~the $W_1$ distance is as follows.
In the preprocessing stage, we build a quadtree $T$ on the ground set $X$, as described in~\Cref{sec:quadtree}.
Let $t(x,x')$ denote the quadtree distance between every pair $x,x'\in X$.
In the query stage, in order to estimate $W_1(\mu,\nu)$ between two distributions $\mu,\nu$, we compute the optimal flow $f$ w.r.t.~the tree metric, that is,
\[
  f = \mathrm{argmin}_{\tilde f}\sum_{x,x'\in X}\tilde f(x,x')\cdot t(x,x') ,
\]
where the argmin is taken over all distributions on $X\times X$ with marginals $\mu,\nu$.
Then, the estimate of the distance between $\mu$ and $\nu$ is given by
\[
  \widetilde{W}_1(\mu,\nu) = \sum_{x,x'\in X}f(x,x')\cdot\norm{x-x'}.
\]

Note that if the support sizes of $\mu$ and $\nu$ are upper-bounded by $s$, then the Flowtree estimate of their distance can be computed in time linear in $s$ (see proof in appendix).

\begin{lemma}\label{lmm:flowtreetime}
$\widetilde{W}_1(\mu,\nu)$ can be computed in time $O(s(d+\log(d\Phi)))$.
\end{lemma}

Unlike Quadtree, Flowtree does not reduce to sparse $\ell_1$ distance computation.
Instead, one needs to compute the optimal flow tree $f$ explicitly by bottom-up greedy algorithm, and then use it to compute $\widetilde{W}_1(\mu, \nu)$. 
On the other hand, Flowtree has the notable property mentioned earlier: its NNS approximation factor is~\emph{independent} of the dataset size $n$. In comparison, the classic Quadtree does not possess this property, and its accuracy deteriorates as the dataset becomes larger.
We formally establish this distinction in two senses: first by analyzing worst-case bounds, and then by analyzing a popular random data model.

\subsection{Worst-case bounds}

We start with an analytic worst-case bound on the performance of quadtree.
Let us recall notation:
$X$ is a finite subset of $\R^d$, and $\Phi>0$ is the side length of a hypercube enclosing $X$.
We are given a dataset of $n$ distributions $\mu_1,\ldots,\mu_n$, and a query distribution $\nu$, where each of these distributions is supported on a subset of $X$ of size at most $s$.
Our goal is to find a near neighbor of $\nu$ among $\mu_1,\ldots\mu_n$. 
A distribution $\mu_i$ is called a~\emph{$c$-approximate nearest neighbor} of $\nu$ if $W_1(\mu_i,\nu)\leq c\cdot\min_{i^*} W_1(\mu_{i^*},\nu)$.

The following theorem is an adaptation of a result by~\cite{andoni2008earth} (where it is proven for a somewhat different algorithm, with similar analysis). All proofs are deferred to the appendix.

\begin{theorem}[Quadtree upper bound]\label{thm:qu}
With probability $\geq0.99$, the nearest neighbor of $\nu$ among $\mu_1,\ldots\mu_n$ in the Quadtree distance is 
an $O(\log(\min\{sn,|X|\})\log(d\Phi))$-approximate nearest neighbor in the $W_1$ distance.
\end{theorem}

Next, we show that the $\log n$ factor in the above upper bound is~\emph{necessary} for Quadtree.

\begin{theorem}[Quadtree lower bound]\label{thm:ql}
Suppose $c$ is such that Quadtree is guaranteed to return a $c$-approximate nearest neighbor, for any dataset, with probability more than (say) $1/2$. Then $c=\Omega(\log n)$.
\end{theorem}

In contrast, Flowtree attains an approximation factor that does not depend on $n$.
\begin{theorem}[Flowtree upper bound]\label{thm:fu}
With probability $\geq0.99$, the nearest neighbor of $\nu$ among $\mu_1,\ldots\mu_n$ in the Flowtree distance is 
an $O(\log(s)\log(d\Phi))$-approximate nearest neighbor for the $W_1$ distance.
\end{theorem}

Finally, we combine ideas from~\cite{andoni2008earth} and~\cite{bavckurs2014better} to prove another upper bound for Flowtree, which is also independent of the dimension $d$ and the numerical range $\Phi$. No such result is known for Quadtree (nor does it follow from our techniques).

\begin{theorem}[Flowtree upper bound for uniform distributions\footnote{For simplicity, \Cref{thm:fu2} is stated for uniform distribution (or close to uniform), such as documents in~\cite{kusner2015word}. A similar result holds for any distribution, with additional dependence on the numerical range of mass values.}]\label{thm:fu2}
For an integer $s$, assume that for every distribution there exists an integer $s'\leq s$ such that the weights of all elements in the support are integer multiples of $1/s'$.
With probability $\geq0.99$, the nearest neighbor of $\nu$ among $\mu_1,\ldots\mu_n$ in the Flowtree distance is an $O(\log^2 s)$-approximate nearest neighbor for the $W_1$ distance.
\end{theorem}

\subsection{Random model}\label{sec:random}
The above worst-case results appear to be overly pessimistic for real data. 
Indeed, in practice we observe that Quadtree and especially Flowtree often recover the exact nearest neighbor. 
This motivates us to study their performance on a simple model of random data, which is standard in the study of nearest neighbor search.

The data is generated as follows. We choose a ground set $X$ of $N$ points i.i.d.~uniformly at random on the $d$-dimensional unit sphere $\mathcal S^{d-1}$.
For each subset of $N$ of size $s$, we form a uniform distribution supported on that subset. These distributions make up the dataset $\mu_1,\ldots,\mu_n$ (so $n={N\choose s}$).

\begin{figure}
\vskip 0.2in
\begin{center}
\includegraphics[width=0.35\textwidth]{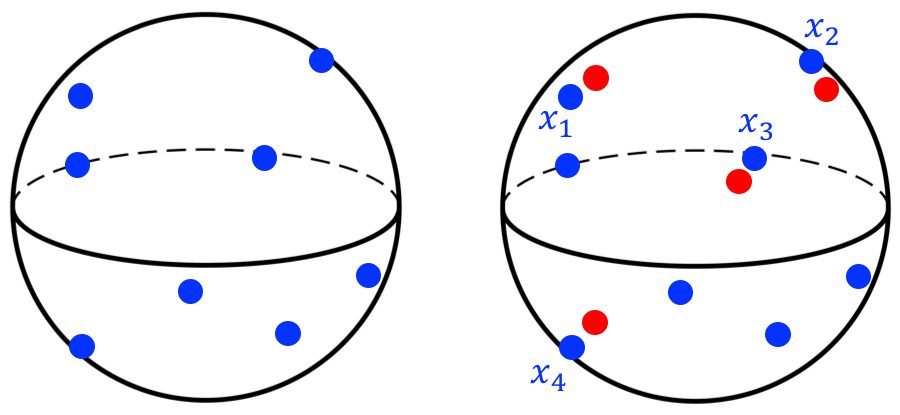}%
\caption{Random model illustration with $s=4$. Left: The blue points are the $N$ random data points. The data distributions are all subsets of $4$ points. Right: The red points form a query distribution whose planted nearest neighbor is the distribution supported on $\{x_1,x_2,x_3,x_4\}$.}
\label{fig:randommodel}
\end{center}
\vskip -0.2in
\end{figure}

To generate a query, pick any $\mu_i$ as the ``planted'' nearest neighbor, and let $x_1,\ldots,x_s$ denote its support.
For $k=1,\ldots,s$, choose a uniformly random point $y_k$ among the points on $\mathcal S^{d-1}$ at distance at most $\epsilon$ from $x_k$, where $\epsilon$ is a model parameter. 
The query distribution $\nu$ is defined as the uniform distribution over $y_1,\ldots,y_s$.
By known concentration of measure results, the distance from $y_k$ to every point in $X$ except $x_k$ is $\sqrt2-o(1)$ with high probability. Thus, the optimal flow from $\nu$ to $\mu_i$ is the perfect matching $\{(x_k,y_k)\}_{k=1}^s$, and $\mu_i$ is the nearest neighbor of $\nu$.
The model is illustrated in Figure~\ref{fig:randommodel}.

\begin{theorem}\label{thm:rm}
In the above model, the success probability of Quadtree in recovering the planted nearest neighbor decays exponentially with $N$, while the success probability of Flowtree is independent of $N$.
\end{theorem}

\section{Experiments}\label{sec:experiments}
In this section we empirically evaluate Flowtree and compare it to various existing methods.

\subsection{Synthetic data}
\begin{figure*}[tbh]
\vskip 0.2in
\begin{center}
\includegraphics[width=0.5\textwidth]{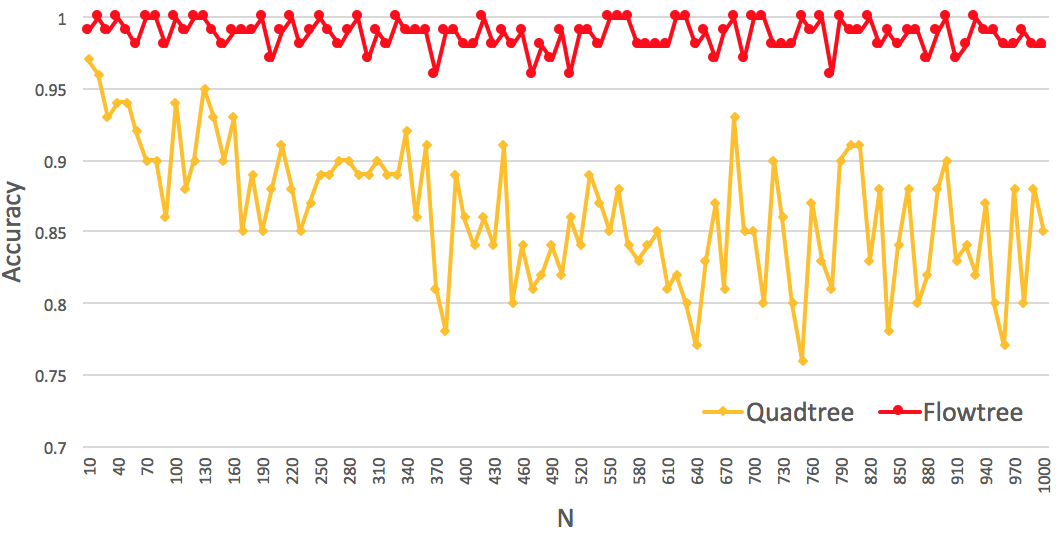}%
\includegraphics[width=0.5\textwidth]{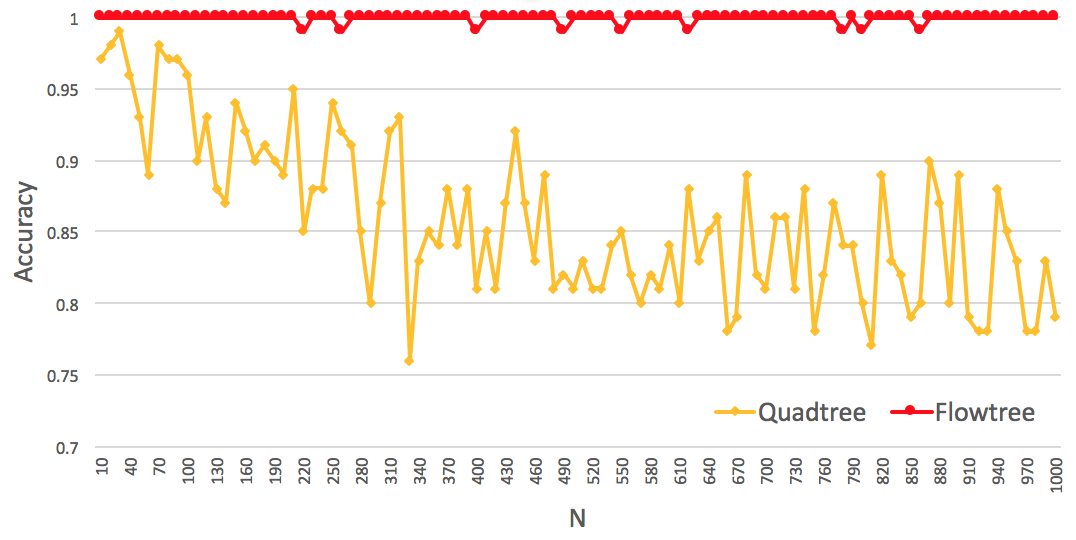}
\caption{Results on random data. Left: $d=s=10$, $\epsilon=0.25$. Right: $d=10$, $s=100$, $\epsilon=0.4$.}
\label{fig:random}
\end{center}
\vskip -0.2in
\end{figure*}

We implement the random model from~\Cref{sec:random}.
The results are in~\Cref{fig:random}.
The x-axis is $N$ (the number of points in the ground metric), and the y-axis is the fraction of successes over $100$ independent repetitions of planting a query and recovering its nearest neighbor. As predicted by~\Cref{thm:rm}, Quadtree's success rate degrades as $N$ increases (and we recall that $n={N\choose s}$), while Flowtree's does not.

\subsection{Real data}

\paragraph{Datasets.}
We use three datasets from two application domains. 
Their properties are summarized in Table~\ref{f:datasets}.

\begin{itemize}
  \item \textit{Text documents:} We use the standard benchmark 20news dataset of news-related online discussion groups, and a dataset of Amazon reviews split evenly over $4$ product categories. Both have been used in~\cite{kusner2015word} to evaluate the Word-Move Distance. Each document is interpreted as a uniform distribution supported on the terms it contains (after stopword removal). For the underlying metric, we use GloVe word embeddings~\cite{pennington2014glove} with $400,000$ terms and $50$ dimensions.
  \item \textit{Image recognition:} We use the MNIST dataset of handwritten digits. As in~\cite{cuturi2013sinkhorn}, each image is interpreted as a distribution over $28\times28$ pixels, with mass proportional to the greyscale intensity of the pixel (normalized so that the mass sums to $1$). Note that the distribution is supported on only the non-white pixels in the image. The underlying metric is the $2$-dimensional Euclidean distance between the $28\times28$ pixels, where they are identified with the points $\{(i,j)\}_{i,j=1}^{28}$ on the plane. 
\end{itemize}

\begin{table}
\centering
\caption{Dataset properties. Avg.~$s$ is the average support size of the distributions in the dataset.}
\label{f:datasets}
\begin{tabular}{llllll}
\toprule
Name & Size & Queries & Underlying metric & Avg.~$s$  \\
\midrule
20news & $11,314$ & $1,000$ & Word embedding & $115.9$   \\
Amazon & $10,000$ & $1,000$ & Word embedding & $57.44$    \\
MNIST & $60,000$ & $10,000$ & 2D Euclidean & $150.07$  \\
\bottomrule
\end{tabular}
\end{table}

\paragraph{Algorithms.}
We evaluate the following algorithms:
\begin{itemize}
  \item\textit{Mean:} $W_1(\mu,\nu)$ is estimated as the Euclidean distance between the means of $\mu$ and $\nu$. This method has been suggested and used in~\cite{kusner2015word}.\footnote{There it is called Word Centroid Distance (WCD).}
  \item\textit{Overlap:} A simple baseline that estimates $W_1(\mu,\nu)$ by the size of the intersection of their supports.
  \item\textit{TF-IDF:} A well-known similarity measure for text documents. It is closely related to Overlap.\footnote{Namely, it is a weighted variant of Overlap, where terms are weighted according to their frequency in the dataset.} For MNIST we omit this baseline since it is not a text dataset.
  \item\textit{Quadtree:} See~\Cref{sec:quadtree}.
  \item\textit{Flowtree:} See~\Cref{sec:flowtree}.
  \item\textit{R-WMD:} The Relaxed WMD method of~\cite{kusner2015word}, described in~\Cref{sec:related}. 
  We remark that this method does not produce an admissible flow (i.e., it does not adhere to the capacity and demand constraints of $W_1$).
  \item\textit{ACT-1}: The Approximate Constrained Transfers method of~\cite{atasu2019linear} gradually adds constraints to R-WMD over $i$ iterations, for a parameter $i$. The $i=0$ case is identical to R-WMD, and increasing $i$ leads to increasing both the accuracy and the running time. Like R-WMD, this method does not produce an admissible flow. In our experiments, the optimal setting for this method is $i=1$,\footnote{This coincides with the results reported in~\cite{atasu2019linear}.} which we denote by ACT-1. The appendix contains additional results for larger $i$.
  \item\textit{Sinkhorn with few iterations:} The iterative Sinkhorn method of~\cite{cuturi2013sinkhorn} is designed to converge to a near-perfect approximation of $W_1$. Nonetheless, it can be adapted into a fast approximation algorithm by invoking it with a fixed small number of iterations. We use 1 and 3 iterations, referred to as Sinkhorn-1 and Sinkhorn-3 respectively. Since the Sinkhorn method requires tuning certain parameters (the number of iterations as well as the regularization parameter), the experiments in this section evaluate the method at its optimal setting, and the appendix includes experiments with other parameter settings.
\end{itemize}

As mentioned in~\Cref{sec:ourresults}, these methods can be grouped by their running time dependence on $s$:
\begin{itemize}
\item \textit{``Fast'' linear-time:} Mean, Overlap, TF-IDF, Quadtree
\item \textit{``Slow'' linear-time:} Flowtree
\item \textit{Quadratic time:} R-WMD, ACT-1, Sinkhorn
\end{itemize}
The difference between ``fast'' and ``slow'' linear time is that the former algorithms reduce to certain simple cache-efficient operations, and furthermore, Mean greatly benefits from SIMD vectorization. 
In particular, Overlap, TF-IDF and Quadtree require computing a single $\ell_1$ distance between sparse vectors, while Mean 
requires computing a single Euclidean distance in the ground metric. 
This renders them an order of magnitude faster than the other methods, as our empirical results will show.

\paragraph{Runtime measurement.}
All running times are measured on a ``Standard F72s\_v2'' Microsoft Azure instance equipped with Intel Xeon Platinum 8168 CPU. In our implementations, we use NumPy linked with OpenBLAS, which is used in a single-threaded mode.

\paragraph{Implementation.}
We implement R-WMD, ACT and Sinkhorn in Python with NumPy, as they amount to standard matrix operations which are handled efficiently by the underlying BLAS implementation. 
We implement Mean, Overlap, TF-IDF, Quadtree and Flowtree in C++ (wrapped in Python for evaluation).
For Mean we use the Eigen library to compute dense $\ell_2$ distances efficiently.
For Exact $W_1$ we use the POT library in Python, which in turn calls the Lemon graph library written in C++.
The accuracy and pipeline evaluation code is in Python.

\begin{figure*}[ht!]

\vskip 0.2in
\begin{center}
\includegraphics[width=0.5\textwidth]{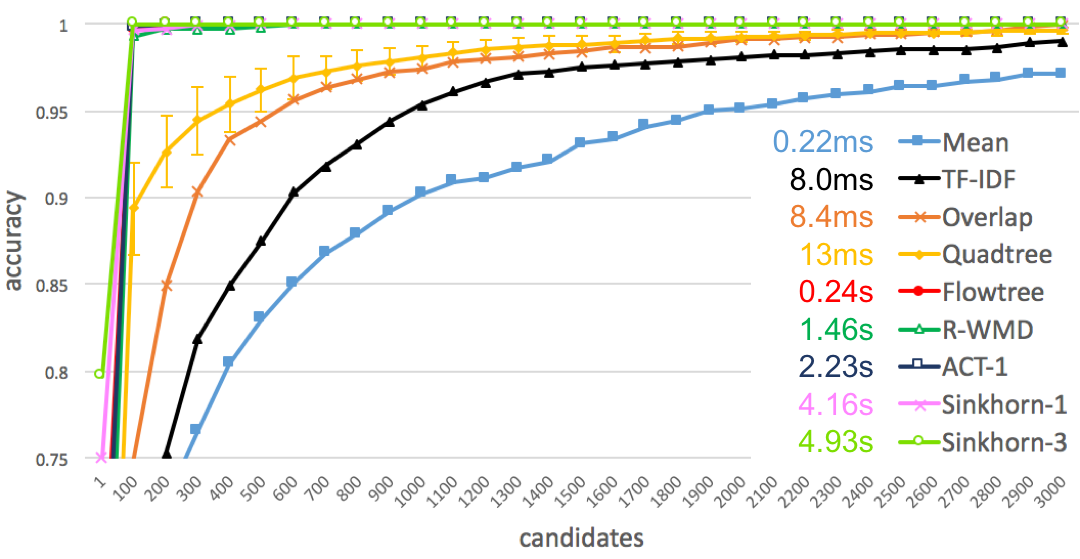}%
\includegraphics[width=0.5\textwidth]{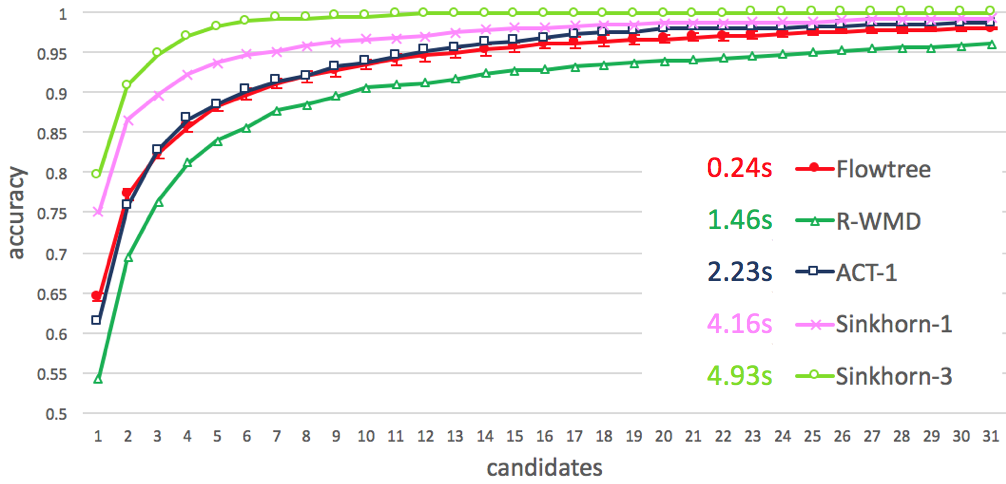}
\caption{Individual accuracy and runtime results on 20news}
\label{fig:20news}
\end{center}

\begin{center}
\includegraphics[width=0.5\textwidth]{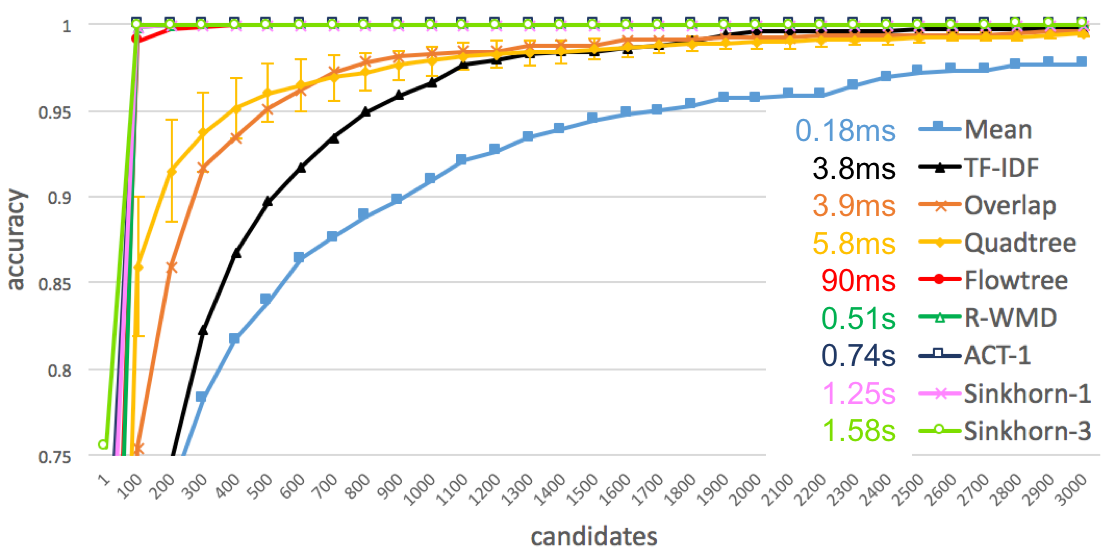}%
\includegraphics[width=0.5\textwidth]{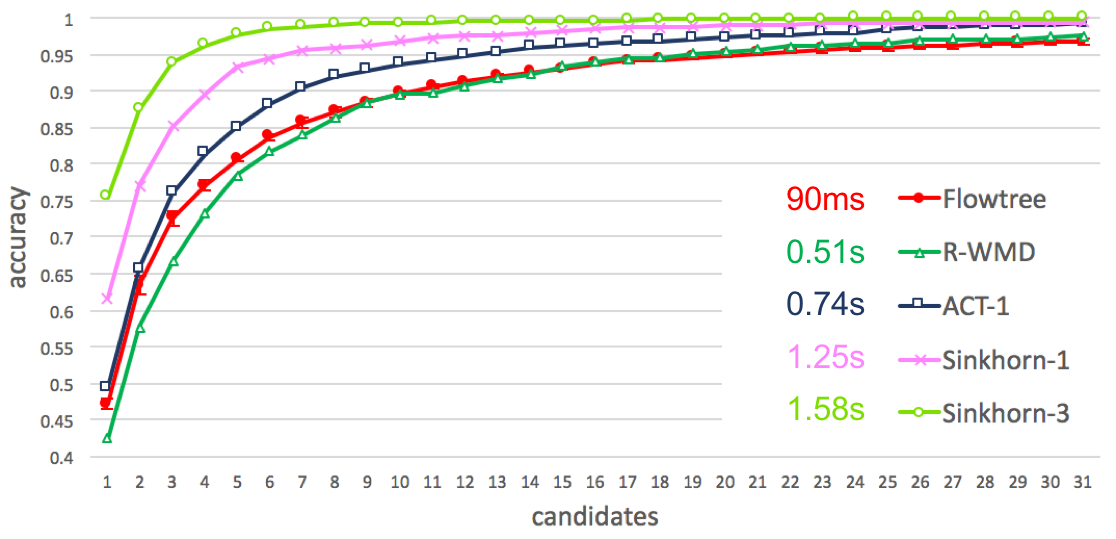}
\caption{Individual accuracy and runtime results on Amazon}
\label{fig:amazon}
\end{center}

\begin{center}
\includegraphics[width=0.5\textwidth]{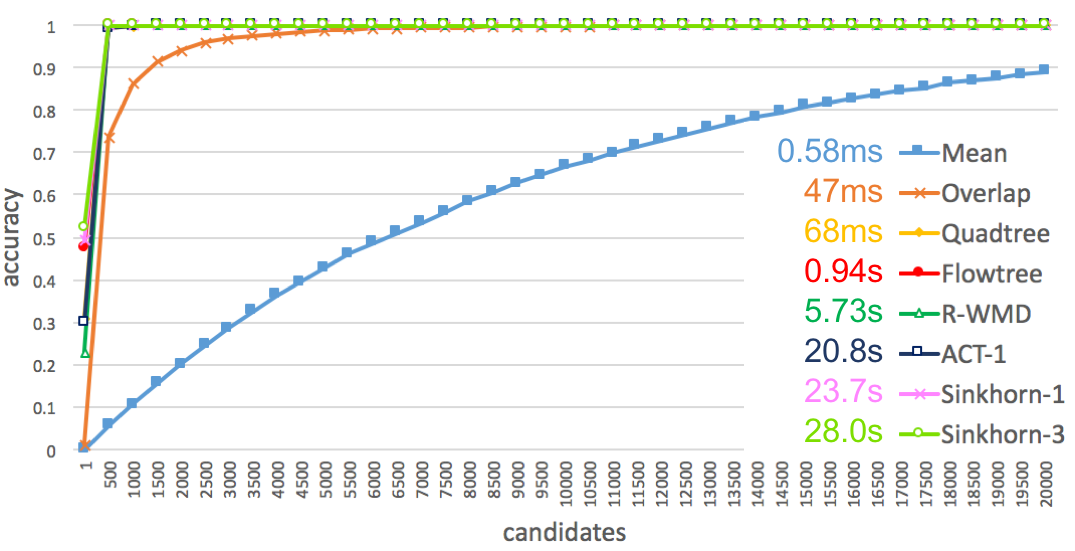}%
\includegraphics[width=0.5\textwidth]{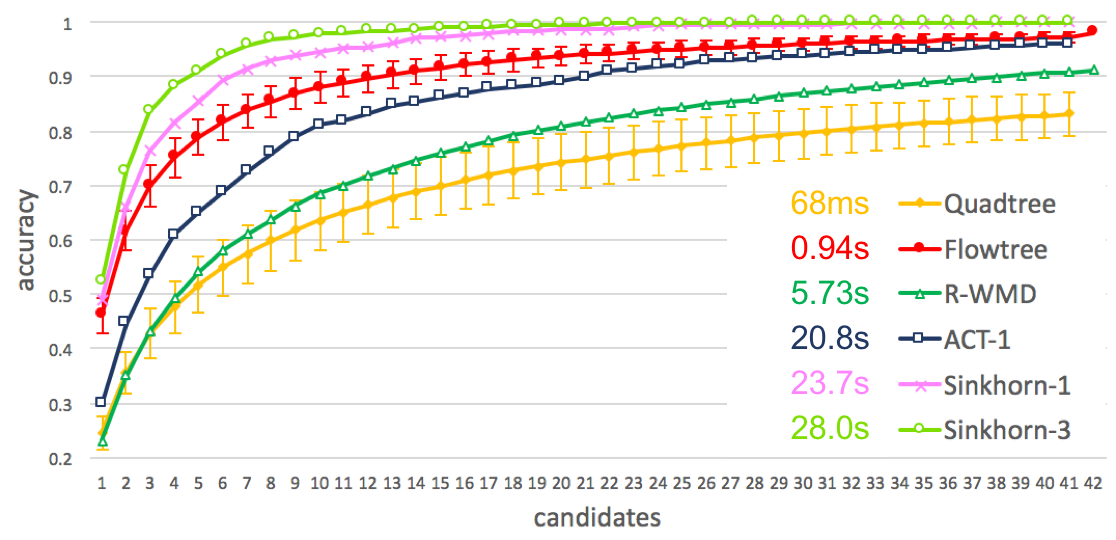}
\caption{Individual accuracy and runtime results on MNIST\textsuperscript{*}}
\label{fig:mnist}
\end{center}
\vskip -0.2in
\end{figure*}
\begin{table*}[h!]
\footnotesize
\caption{Running times}
\label{f:runtimes}
\medskip
\centering
\begin{tabular}{lllllllllll}
\toprule
Dataset & Mean & TF-IDF & Overlap & Quadtree & Flowtree & R-WMD & ACT-1\textsuperscript{**} & Sinkhorn-1 & Sinkhorn-3 & Exact $W_1$  \\
\midrule
20news & $0.22$ms & $8.0$ms & $8.4$ms & $13$ms & $0.24$s & $1.46$s & $2.23$s &$4.16$s & $4.93$s & $41.5$s \\
Amazon & $0.18$ms & $3.8$ms & $3.9$ms & $5.8$ms & $90$ms & $0.51$s & $0.74$s & $1.25$s  & $1.58$s  & $4.23$s  \\
MNIST\textsuperscript{*} & $0.58$ms & --- & $47$ms & $68$ms & $0.94$s & $5.73$s & $20.8$s & $23.7$s & $28.0$s & $154.0$s \\
\bottomrule
\end{tabular}
\begin{flushleft}
\begin{footnotesize}
\smallskip
\textsuperscript{*}  On MNIST, the accuracy of R-WMD, ACT and Sinkhorn is evaluated on $1,000$ random queries. The running time of R-WMD, ACT, Sinkhorn and Exact $W_1$ is measured on $100$ random queries. The running time of Flowtree is measured $1,000$ random queries.\\
\textsuperscript{**}  ACT takes a faster form when applied to uniform distributions. We use a separate implementation for this case. This accounts for the large difference in its performance on 20news and Amazon (where distributions are uniform) compared to MNIST (where they are not).
\end{footnotesize}
\end{flushleft}
\end{table*}

\subsection{Individual accuracy experiments}

Our first set of experiments evaluates the runtime and accuracy of each algorithm individually. 
The results are depicted in~\Cref{fig:amazon,fig:20news,fig:mnist}. 
The plots report the recall@$m$ accuracy as $m$ grows. 
The recall@$m$ accuracy is defined as the fraction of queries for which the true nearest neighbors is included in the top-$m$ ranked neighbors (called~\emph{candidates}) by the evaluated method.

For each dataset, the left plot reports the accuracy of all of the methods for large values of $m$.
The right plot reports the accuracy of the high-accuracy methods for smaller values of $m$ (since they cannot be discerned in the left plots).
The high-accuracy methods are Flowtree, R-WMD, ACT, Sinkhorn, and on MNIST also Quadtree.
For Quadtree and Flowtree, which are randomized methods, we report the mean and standard deviation (shown as error bars) of $5$ executions. The other methods are deterministic. 
The legend of each plot is annonated with the running time of each method, also summarized in Table~\ref{f:runtimes}.

\paragraph{Results.}
The tested algorithms yield a wide spectrum of different time-accuracy tradeoffs. 
The ``fast'' linear time methods (Mean, TF-IDF, Overlap and Quadtree) run in order of milliseconds, but are less accurate than the rest. 
The quadratic time methods (R-WMD, ACT-1 and Sinkhorn) are much slower, running in order of seconds, but are dramatically more accurate.

Flowtree achieves comparable accuracy to the quadratic time baselines, while being faster by a margin.
In particular, its accuracy is either similar to or better than R-WMD, while being $5.5$ to $6$ times faster.
Compared to ACT-1, Flowtree is either somewhat less or more accurate (depending on the dataset), while being at least $8$ times faster.
Compared to Sinkhorn, Flowtree achieves somewhat lower accuracy, but is at least $13.8$ and up to $30$ times faster.

\subsection{Pipeline experiments}

The above results exhibit a sharp divide between fast and coarse algorithms to slow and accurate ones.
In practical nearest neighbor search system, both types of algorithms are often combined sequentially as a~\emph{pipeline} (e.g., \cite{sivic2003video,jegou2008hamming,jegou2010product}). 
First, a fast and coarse method is applied to all points, pruning most of them; then a slower and more accurate method is applied to the surviving points, pruning them further; and so on, until finally exact computation is performed on a small number of surviving points. 
In particular, \cite{kusner2015word} employ such a pipeline for the Word Mover Distance, which combines Mean, R-WMD, and exact $W_1$ computation.

In this section, we systematically evaluate pipelines built of the algorithms tested above, on the 20news dataset.

\paragraph{Experimental setup.}
We perform two sets of experiments: In one, the pipeline reports one candidate, and its goal is to output the true nearest neighbor (i.e., recall@1). 
In the other, the pipeline reports $5$ candidates, and its goal is to include the true nearest neighbor among them (i.e., recall@5). 
We fix the target accuracy to $0.9$ (i.e., the pipeline must achieve the recall goal on $90\%$ of the queries), and report its median running time over 3 identical runs. 

\paragraph{Evaluated pipelines.}
The baseline pipelines we consider contain up to three methods:
\begin{itemize}
  \item First: Mean, Overlap or Quadtree.
  \item Second: R-WMD, ACT-1, Sinkhorn-1, or Sinkhorn-3.
  \item Third: Exact $W_1$ computation. For recall@5 pipelines whose second method is Sinkhorn, this third step is omitted, since they already attain the accuracy goal without it. 
\end{itemize}
To introduce Flowtree into the pipelines, we evaluate it both as an intermediate stage between the first and second methods, and as a replacement for the second method.

\paragraph{Pipeline parameters.}
A pipeline with $\ell$ algorithms has parameters $c_1,\ldots,c_{\ell-1}$, where $c_i$ is the number of output candidates (non-pruned points) of the $i^{th}$ algorithm in the pipeline.\footnote{The final algorithm always outputs either $1$ or $5$ points, according to the recall goal.} 
We tune the parameters of each pipeline optimally on a random subset of $300$ queries (fixed for all pipelines).
The optimal parameters are listed in the appendix.

\begin{figure*}[t]
\vskip 0.2in
\begin{center}
\includegraphics[width=0.5\textwidth]{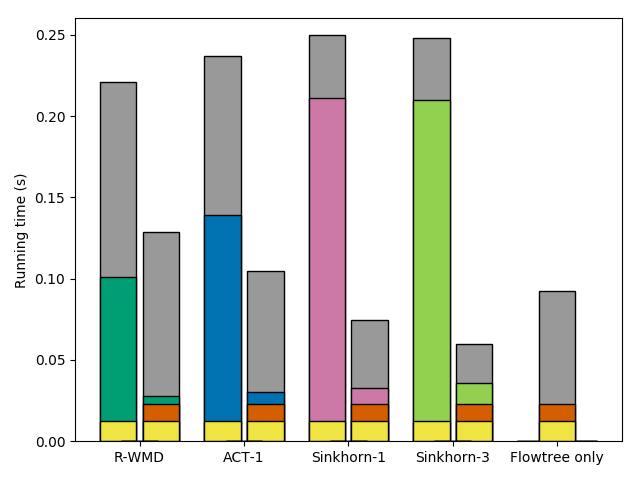}%
\includegraphics[width=0.5\textwidth]{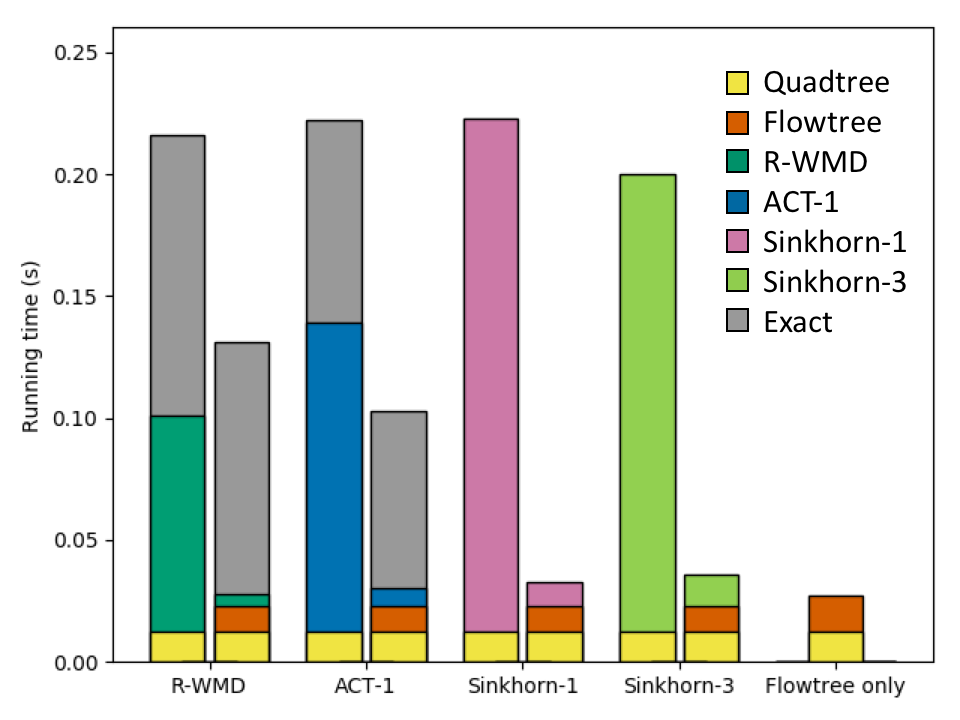}
\caption{Best performing pipelines for recall@1 $\geq0.9$ (on the left) and recall@5 $\geq0.9$ (on the right). Each vertical bar denotes a pipeline, built of the methods indicated by the color encoding, bottom-up. The y-axis measures the running time up to each step of the pipeline. The plot depicts 4 baseline pipelines, consisting of Quadtree, then the method $X$ indicated on the x-axis (R-WMD, ACT-1, Sinkhorn-1 or Sinkhorn-3), and then (optionally) Exact $W_1$. Next to each baseline bar we show the bar obtained by adding Flowtree to the pipeline as an intermediate algorithm between Quadtree and $X$. 
The rightmost bar in each plot shows the pipeline obtained by using Flowtree instead of $X$.}
\label{fig:pipeline}
\end{center}
\vskip -0.2in
\end{figure*}

\begin{table}
\centering
\caption{Best pipeline runtime results}
\label{f:pipeline}
\begin{tabular}{lcc}
\toprule
 & Recall@1 $\geq0.9$ & Recall@5 $\geq0.9$  \\
\midrule
Without Flowtree & 0.221s & 0.200s   \\
With Flowtree & 0.059s & 0.027s    \\
\bottomrule
\end{tabular}
\end{table}

\paragraph{Results.} We found that in the first step of the pipeline, Quadtree is significantly preferable to Mean and Overlap, and the results reported in this section are restricted to it. More results are included in the appendix.

Figure~\ref{fig:pipeline} shows the runtimes of pipelines that start with Quadtree. 
Note that each pipeline is optimized by different parameters, not depicted in the figure. For example, Sinkhorn-3 is faster than Sinkhorn-1 on the right plot, even though it is generally a slower algorithm. However, it is also more accurate, which allows the preceding Quadtree step to be less accurate and report fewer candidates, while still attaining overall accuracy of $0.9$. Specifically, in the optimal recall@5 setting, Sinkhorn-3 runs on 227 candidates reported by Quadtree, while Sinkhorn-1 runs on 295.

The best runtimes are summarized in Table~\ref{f:pipeline}. 
The results show that introducing Flowtree improves the best runtimes by a factor of $3.7$ for recall@1 pipelines, and by a factor of $7.4$ for recall@5 pipelines.

In the recall@1 experiments, the optimally tuned baseline pipelines attain runtimes between $0.22$ to $0.25$ seconds. 
Introducing Flowtree before the second method in each pipeline improves its running time by a factor of $1.7$ to $4.15$. Introducing Flowtree instead of the second method improves the runtime by a factor of $2.4$ to $2.7$.

Once Flowtree is introduced into the recall@1 pipelines, the primary bottleneck becomes the final stage of exact $W_1$ computations. In the recall@5 experiments, this step is not always required, which enables larger gains for Flowtree. 
In these experiments, the optimally tuned baseline pipelines attain runtimes between $0.2$ to $0.22$ seconds. 
Introducing Flowtree before the second method in each pipeline improves its running time by a factor of $1.64$ to $6.75$. Introducing Flowtree instead of the second method improves the runtime by a factor of $7.4$ to $8$.

Overall, Flowtree significantly improves the running time of every pipeline, both as an addition and as a replacement.

\paragraph{Acknowledgments.}
Supported by NSF TRIPODS awards No.~1740751 and No.~1535851, Simons Investigator Award, and MIT-IBM Watson AI Lab collaboration grant.

\bibliographystyle{alpha}
\bibliography{bib}

\appendix

\section{Proofs}

\subsection{Flowtree Computation}
In this section we prove~\Cref{lmm:flowtreetime}. We begin by specifying the greedy flow computation algorithm on the tree. Let $h$ denote the height of the tree (for the quadtree the height is $h=O(\log(d\Phi))$. 
Suppose we are given a pair of distributions $\mu,\nu$, each supported on at most $s$ leaves of the tree. For every node $v$ in the tree, let $C_\mu(v)$ denote the set of points in $x\in X$ such that $\mu(x)>0$ and the tree leaf that contains $x$ is a descendant of $v$. Similarly define $C_\nu(v)$. Note that we only need to consider nodes for which either $C_\mu(v)$ or $C_\nu(v)$ is non-empty, and there are at most $2sh$ such nodes.

The algorithm starts with a zero flow $f$, and processes the nodes in a bottom-up order starting at the leaf. In each node, the unmatched demands collected from its children are matched arbitrarily, and the demands that cannot be matched are passed on to the parent. In mode detail, a node is processed as follows:
\begin{enumerate}
  \item Collect from the children the list of unmatched $\mu$-demands for the nodes in $C_\mu(v)$ and the list of unmatched $\nu$-demands for the nodes in $C_\nu(v)$. Let $\{\mu_v(x):x\in C_\mu(v)\}$ denote the unmatched $\nu$-demands and let $\{\nu_v(x):x\in C_\nu(v)\}$ denote the unmatched $\nu$-demands.
  \item While there is a pair $x\in C_\mu(v)$ and $x'\in C_\mu(v)$ with $\mu_v(x)>0$ and $\nu_v(x')>0$, let $\eta=\min\{\mu_v(x),\nu_v(x')\}$, and update (i) $f(x,x')\mathrel{{+}{=}}\eta$, (ii) $\mu_v(x)\mathrel{{-}{=}}\eta$, (iii) $\nu_v(x')\mathrel{{-}{=}}\eta$.
  \item Now either $\mu_v$ or $\nu_v$ is all-zeros. If the other one is not all-zeros (i.e., there is either remaining unmatched $\mu$-demand or remaining unmatched $\nu$-demand), pass it on the the parent.
\end{enumerate}
A leaf $v$ contains a single point $x\in X$ with either $\mu(x)>0$ or $\nu(x)>0$; it simply passes it on to its parent without processing.

It is well known that the above algorithm computes an optimal flow on the tree (with respect to tree distance costs), see, e.g., \cite{kalantari1995linear}. 
Let us now bound its running time.
The processing time per node $v$ in the above algorithm is $O(|C_\mu(v)|+|C_\nu(v)|)$. 
In every given level in the tree, if $v_1,\ldots,v_k$ are the nodes in that level, then $\{C_\mu(v_1),\ldots,C_\mu(v_k)\}$ is a partition of the support of $\mu$, and $\{C_\nu(v_1),\ldots,C_\nu(v_k)\}$ is a partition of the support of $\nu$. Therefore the total processing time per level is $O(s)$, and since there are $h$ levels, the flow computation time is $O(sh)$.
Then we need to compute the Flowtree output $\widetilde{W}_1(\mu,\nu)$. 
Observe that in the above algorithm, whenever we match demands between a pair $x,x'$, we fully satisfy the unmatched demand of one of them. Therefore the output flow $f$ puts non-zero flow between at most $2s$ pairs. For each such pair we need to compute the Euclidean distance in time $O(d)$, and the overall running time is $O(s(d+h))$.

\subsection{Quadtree and Flowtree Analysis}

\begin{proof}[Proof of~\Cref{thm:qu}]
Let $x,y\in X$.
Let $p_\ell(x,y)$ be the probability that $x,y$ fall into the same cell (hypercube) in level $\ell$ of the quadtree. It is not hard to see that it satisfies,
\[
  1-\frac{\norm{x-y}_1}{2^\ell} \leq
  p_\ell(x,y) \leq
  \exp\left(-\frac{\norm{x-y}_1}{2^\ell}\right).
\]
Let $t$ be the tree metric induced on $X$ by the quadtree.
Note that for $t(x,y)$ to be at most $O(2^\ell)$, $x,y$ must fall into the same hypercube in level $\ell$.
For any $\delta>0$, we can round $\frac{\norm{x-y}_1}{\log(1/\delta)}$ to its nearest power of $2$ and obtain $\ell$ such that $2^\ell=\Theta(\frac{\norm{x-y}_1}{\log(1/\delta)})$. It satisfies,
\[
  \Pr\left[t(x,y) < \frac{O(1)}{\log(1/\delta)}\norm{x-y}_1\right] \leq\delta.
\]
By letting $\delta=\Omega(\min\{1/|X|,1/(s^2n)\})$, we can take union bound either over all pairwise distances in $X$ (of which there are ${|X|\choose2}$), or over all distances between the support of the query $\nu$ and the union of supports of the dataset $\mu_1,\ldots,\mu_n$ (of which there are at most $s^2n$, if every support has size at most $s$).
Then, with probability say $0.995$, all those distances are contracted by at most $O(\log(\min\{sn,|X|\}))$, i.e.,
\begin{equation}\label{eq:qcontract}
t(x,y) \geq \frac{1}{O(\log(1/\delta))}\norm{x-y}_1.
\end{equation}

On the other hand,

\[
  \E[t(x,y)] = \sum_\ell2^\ell\cdot(1-p_\ell(x,y)) \leq \sum_{\ell}2^\ell\cdot\frac{\norm{x-y}_1}{2^\ell} \leq
  O(\log(d\Phi))\cdot\norm{x-y}_1.
\]
Let $\mu^*$ be the true nearest neighbor of $\nu$ in $\mu_1,\ldots,\mu_n$. Let $f^*_{\mu^*,\nu}$ be the optimal flow between them.
Then by the above,

\[
  \E\left[\sum_{(x,y)\in X\times X}f^*_{\mu^*,\nu}(x,y)t(x,y)\right] \leq O(\log(d\Phi))\sum_{(x,y)\in X\times X}f^*_{\mu^*,\nu}(x,y)\norm{x-y}_1 .
\]
By Markov, with probability say $0.995$,
\begin{equation}\label{eq:qexpand}
 \sum_{(x,y)\in X\times X}f^*_{\mu^*,\nu}(x,y)\cdot t(x,y) \leq O(\log(d\Phi))\sum_{(x,y)\in X\times X}f^*_{\mu^*,\nu}(x,y)\cdot\norm{x-y}_1 .
\end{equation}
Let $\mu'$ be the nearest neighbor of $\nu$ in the dataset according to the quadtree distance.
Let $f^*_{\mu',\nu}$ be the optimal flow between them in the true underlying metric ($\ell_1$ on $X$), and let $f_{\mu,\nu}$ be the optimal flow in the quadtree. Finally let $W_t$ denote the Wasserstein-$1$ distance on the quadtree. Then,
\begin{align*}
  & W_1(\mu',\nu)\\
  &= \sum_{(x,y)\in X\times X}f^*_{\mu',\nu}(x,y)\cdot\norm{x-y}_1 & \\
  &\leq \sum_{(x,y)\in X\times X}f_{\mu',\nu}\cdot\norm{x-y}_1 &
\text{$f^*_{\mu^*,\nu}$ is optimal for $\norm{\cdot}_1$} \\ 
  &\leq O(\log(\min\{sn,|X|\}))\sum_{(x,x')\in X\times X}f_{\mu',\nu}\cdot t(x,y) &
\text{\cref{eq:qcontract}} \\  
  &= O(\log(\min\{sn,|X|\}))\cdot W_t(\mu',\nu) & \text{definition of $W_t$} \\
  &\leq O(\log(\min\{sn,|X|\}))\cdot W_t(\mu^*,\nu) & \text{$\mu'$ is the nearest neighbor in $W_t$} \\
  &= O(\log(\min\{sn,|X|\}))\sum_{(x,y)\in X\times X}f_{\mu^*,\nu}\cdot t(x,y) &
\text{definition of $W_t$} \\  
  &\leq O(\log(\min\{sn,|X|\}))\sum_{(x,y)\in X\times X}f^*_{\mu^*,\nu}\cdot t(x,y) & \text{$f_{\mu^*,\nu}$ is optimal for $t(\cdot,\cdot)$} \\  & \leq O(\log(\min\{sn,|X|\})\log(d\Phi))\sum_{(x,y)\in X\times X}f^*_{\mu^*,\nu}\cdot \norm{x-y}_1 & \text{\cref{eq:qexpand}} \\
  &= O(\log(\min\{sn,|X|\})\log(d\Phi)) \cdot W_1(\mu^*,\nu) ,
\end{align*}

so $\mu'$ is a $O(\log(\min\{sn,|X|\})\log(d\Phi))$-approximate nearest neighbor.
\end{proof}

\begin{proof}[Proof of~\Cref{thm:ql}]
It suffices to prove the claim for $s=1$ (i.e., the standard $\ell_1$-distance).
Let $d>0$ be an even integer.
Consider the $d$-dimensional hypercube.
Our query point is the origin.
The true nearest neighbor is $e_1$ (standard basis vector).
The other data points are the hypercube nodes whose hamming weight is exactly $d/2$. The number of such points is $\Theta(2^d/\sqrt{d})$, and this is our $n$.

Consider imposing the grid with cell side $2$ on the hypercube. The probability that $0$ and $1$ are uncut in a given axis is exactly $1/2$, and since the shifts in different axes are independent, the number of uncut axes is distributed as $Bin(d,1/2)$.
Thus with probability $1/2$ there are at least $d/2$ uncut dimensions. If this happens, we have a data point hashed into the same grid cell as the origin (to get such data point, put $1$ in any $d/2$ uncut dimensions and $0$ in the rest), so its quadtree distance from the origin is $1$.
On the other hand, the distance of the origin to its true nearest neighbor $e_1$ is at least $1$, since they will necessarily be separated in the next level (when the grid cells have side $1$). Thus the quadtree cannot tell between the true nearest neighbor and the one at distance $d/2$, and we get the lower bound $c\geq d/2$. Since $n=\Theta(2^d/\sqrt{d})$, we have $d/2=\Omega(\log n)$ as desired.
\end{proof}

\begin{proof}[Proof of~\Cref{thm:fu}]
The proof is the same as for~\Cref{thm:qu}, except that in~\cref{eq:qcontract}, we take a union bound only over the $s^2$ distances between the supports of $\nu$ and $\mu^*$ (the query and its true nearest neighbor).
Thus each distance between $\mu^*$ and $\nu$ is contracted by at most $O(\log s)$.

Let $W_F$ denote the Flowtree distance estimate of $W_1$.
Let $\mu'$ be the nearest neighbor of $\nu$ in the Flowtree distance.
With the same notation in the proof of~\Cref{thm:qu},
\begin{align*}
    W_1(\mu',\nu) &= \sum_{(x,y)\in X\times X}f^*_{\mu',\nu}(x,y)\cdot\norm{x-y}_1  &  \\
    &\leq \sum_{(x,y)\times X\times X}f_{\mu',\nu}(x,y)\cdot\norm{x-y}_1 & \text{$f^*_{\mu',\nu}$ is optimal for $\norm{\cdot}_1$} \\
    &= W_F(\mu',\nu) & \text{Flowtree definition} \\
    &\leq W_F(\mu^*,\nu) &  \text{$\mu'$ is nearest in Flowtree distance} \\
    &= \sum_{(x,y)\in X\times X}f_{\mu^*,\nu}(x,y)\cdot\norm{x-y}_1 & \text{Flowtree definition} \\
    &\leq O(\log s) \sum_{(x,y)\in X\times X}f_{\mu^*,\nu}(x,y)\cdot t(x,y) & \text{\cref{eq:qcontract}} \\
    &\leq O(\log s) \sum_{(x,y)\in X\times X}f^*_{\mu^*,\nu}(x,y)\cdot t(x,y) & \text{$f_{\mu^*,\nu}$ is optimal for $t(\cdot,\cdot)$} \\
    &\leq O(\log(d\Phi)\log s) \sum_{(x,y)\in X\times X}f^*_{\mu^*,\nu}(x,y)\cdot\norm{x-y}_1 & \text{\cref{eq:qexpand}} \\
    & = O(\log(d\Phi)\log s) \cdot W_1(\mu^*,\nu),
\end{align*}
as needed. Note that the difference from the proof of~\Cref{thm:qu} is that we only needed the contraction bound (\cref{eq:qcontract}) for distances between $\mu^*$ and $\nu$.
\end{proof}

\begin{proof}[Proof of~\Cref{thm:fu2}]

We set $\eps = 1/\log s$. Let $t'(x,y)$ denote the quadtree distance where the weight corresponding to a cell $v$ in level $\ell(v)$ is $2^{\ell(v)(1-\eps)}$ instead of $2^{\ell(v)}$. Let $f_{\mu,\nu}$ be the optimal flow in the quadtree defined by weights $t'$.

Let $\delta=c/s^2$ where $c>0$ is a sufficiently small constant.
For a every $x,y$, let $\ell_{xy}$ be the largest integer such that
\[ 2^{\ell_{xy}}\leq \frac{\norm{x-y}_1}{(\log(1/\delta))^{1/(1-\epsilon)}} . \]
The probability that $x,y$ are separated (i.e., they are in different quadtree cells) in level $\ell_{xy}$ is
\[
  1-p_{\ell_{xy}(x,y)} \geq 1-\exp\left(-\frac{\norm{x-y}_1}{2^{\ell_{xy}}}\right) \geq 1 -\frac{\delta}{1-\epsilon} .
\]
By the setting of $\delta$, we can take a union bound over all $x\in\mathrm{support(\mu^*)}$ and $y\in\mathrm{support(\nu)}$ and obtain that with say $0.99$ probability, simultaneously, every pair $x,y$ is separated at level $\ell_{xy}$.
We denote this event by $\mathcal E_{lower}$ and suppose it occurs.
Then for every $x,y$ we have
\begin{equation} \label{eq:qcontract2}
  t'(x,y) \geq 2\cdot2^{\ell_{xy}(1-\epsilon)} \nonumber \geq 2\cdot\left(\frac12\cdot\frac{\norm{x-y}_1}{(\log(1/\delta))^{1/(1-\epsilon)}}\right)^{1-\epsilon}  \geq
  \frac{\norm{x-y}_1^{1-\epsilon}}{\log(1/\delta)} =
  \frac{\norm{x-y}_1^{1-\epsilon}}{\Theta(\log s)} .
\end{equation}


Next we upper-bound the expected tree distance $t'(x,y)$. (Note that we are not conditioning on $\mathcal E_{lower}$.)
Observe that
\[ t'(x,y) = 2\sum_{\ell=-\infty}^\infty2^{\ell(1-\epsilon)}\cdot\mathbf1\{\text{$x,y$ are separated at level $\ell$}\} . \]
Let $L_{x,y}$ be the largest integer such that $2^{L_{xy}}\leq \norm{x-y}_1$.
We break up $t'(x,y)$ into two terms,
\[
t_{lower}'(x,y) =2\sum_{\ell=-\infty}^{L_{xy}}2^{\ell(1-\epsilon)}\cdot\mathbf1\{\text{$x,y$ are separated at level $\ell$}\} , 
\]
and
\[ t_{upper}'(x,y) =2\sum_{\ell=L_{xy}+1}^{\infty}2^{\ell(1-\epsilon)}\cdot\mathbf1\{\text{$x,y$ are separated at level $\ell$}\} , 
\]
thus $t'(x,y)=t_{lower}'(x,y)+t_{upper}'(x,y)$.
For $t_{lower}'(x,y)$ it is clear that deterministically,
\[
  t_{lower}'(x,y) \leq 2\sum_{\ell=-\infty}^{L_{xy}}2^{\ell(1-\epsilon)} = O\left(2^{L_{xy}(1-\epsilon)}\right) =
  O\left(\norm{x-y}_1^{1-\epsilon}\right) .
\]
For $t_{upper}'(x,y)$, we have

\begin{align*}
  \E[t_{upper}'(x,y)] &= 2\sum_{\ell=L_{xy}+1}^{\infty}2^{\ell(1-\epsilon)}p_\ell(x,y) \\
  & \leq 2\sum_{\ell=L_{xy}}^{\infty}2^{\ell(1-\epsilon)}\cdot\frac{\norm{x-y}_1}{2^\ell} \\
  &= 2\norm{x-y}_1\sum_{\ell=L_{xy}}^{\infty}2^{-\epsilon\ell} \\
  &= 2\norm{x-y}_1\cdot\frac{2^{-L_{xy}\cdot\epsilon}}{1-2^{-\epsilon}} \\
  &\leq O(\log s)\cdot\norm{x-y}_1^{1-\epsilon} ,
\end{align*}
where in the final bound we have used that $2^{L_{xy}}=\Theta(\norm{x-y}_1)$ and $1-2^{-\epsilon}=\Theta(\epsilon)=\Theta(\log s)$.
Together,
\begin{equation} \label{eq1}
  \E[t'(x,y)]  = \E[t_{lower}'(x,y)+t_{upper}'(x,y)]  \leq \Theta(\log s) \cdot \norm{x-y}_1^{1-\epsilon} .
\end{equation}

Now we are ready to show the $O(\log^2 s)$ upper bound on the approximation factor.
Below we will use the fact that every weight $f_{\mu^*,\nu}(x,y)$ in the flow is of the form $i/(s's'')$ for some integer $0\leq i \leq s's''$. This follows from the assumption that each element in the support of every measure is an integer multiple of $1/s'$ or of $1/s''$ for some $1\leq s',s''\leq s$.
\begin{align*}
    W_1(\mu',\nu) &= \sum_{(x,y)\in X\times X}f^*_{\mu',\nu}(x,y)\cdot\norm{x-y}_1 &  \text{$f^*_{\mu',\nu}$ is optimal for $\norm{\cdot}_1$} \\
    &\leq \sum_{(x,y)\times X\times X}f_{\mu',\nu}(x,y)\cdot\norm{x-y}_1 & \\
    &= W_F(\mu',\nu) & \text{Flowtree definition} \\
    &\leq W_F(\mu^*,\nu) & \text{$\mu'$ is nearest to $\nu$ in Flowtree distance} \\
    &= \sum_{(x,y)\in X\times X}f_{\mu^*,\nu}(x,y)\cdot\norm{x-y}_1 & \text{Flowtree definition} \\
	& \leq \left(\sum_{(x,y)\in X\times X}f_{\mu^*,\nu}^{1-\eps}(x,y)\cdot\norm{x-y}_1^{1-\eps}\right)^{1/(1-\eps)}  &  \text{subadditivity of $(\cdot)^{1-\eps}$} \\
	& \leq O(1)\left(\sum_{(x,y)\in X\times X}f_{\mu^*,\nu}(x,y)\cdot\norm{x-y}_1^{1-\eps}\right)^{1/(1-\eps)}  &  {f_{\mu^*, \nu}(x,y) \geq 1/(s's'') \geq 1/s^2 \text{ or } f_{\mu^*, \nu}(x,y) = 0} \\
    &\leq O(\log s) \left(\sum_{(x,y)\in X\times X}f_{\mu^*,\nu}(x,y)\cdot t'(x,y)\right)^{1/(1-\eps)} &  \text{\cref{eq:qcontract2}} \\
    &\leq O(\log s) \left( \sum_{(x,y)\in X\times X}f^*_{\mu^*,\nu}(x,y)\cdot t'(x,y) \right)^{1/(1-\eps)} & \text{$f_{\mu^*,\nu}$ is optimal for $t'(\cdot,\cdot)$} \\
    &\leq O(\log^2 s) \left(\sum_{(x,y)\in X\times X}f^*_{\mu^*,\nu}(x,y)\cdot\norm{x-y}_1^{1-\eps} \right)^{1/(1-\eps)} &\text{\cref{eq1}} \\
	&\leq O(\log^2 s) \sum_{(x,y)\in X\times X}f^*_{\mu^*,\nu}(x,y)\cdot\norm{x-y}_1 & \text{concavity of $(\cdot)^{1-\eps}$ and $\sum f^*_{\mu^*,\nu}(x,y)=1$} \\
    & \leq O(\log^2 s) \cdot W_1(\mu^*,\nu),
\end{align*}
as needed.
\end{proof}

\begin{proof}[Proof of~\Cref{thm:rm}]
{\bf Quadtree.}
For every $k=1,\ldots,s$, let $H_k$ be the smallest hypercube in the quadtree that contains both $x_k$ and $y_k$.
(Note that $H_k$ is a random variable, determined by the initial random shift in the Quadtree construction.)
In order for Quadtree to correctly identify $\mu_i$ as the nearest neighbor of $\nu$, every $H_k$ must not contain any additional points from $X$. Otherwise, if say $H_1$ contains a point $x'\neq x_1$, the $W_1$ distance on the quadtree from $\nu$ to $\mu_i$ is equal to its distance to the uniform distribution over $\{x',x_2,\ldots,x_s\}$.
Since the points in $X$ are chosen uniformly i.i.d.~over $\mathcal S^{d-1}$, the probability of the above event, and thus the success probability of Quadtree, is upper bounded by $\E[(1-V)^{N-s}]$, where $V=\mathrm{volume}(\cup_{k=1}^s H_k\cap\mathcal S^{d-1})$. This $V$ is a random variable whose distribution depends only on $d,s,\epsilon$, and is independent of $N$. Thus the success probability decays exponentially with $N$.

{\bf Flowtree.}
On the other hand, suppose that each $H_k$ contains no other points from $\{x_1,\ldots,x_s\}$ other than $x_k$ (but is allowed to contain any other points from $X$).
This event guarantees that the optimal flow on the tree between $\mu_i$ and $\nu$ is the planted perfect matching, i.e., the true optimal flow, and thus the estimated Flowtree distance between them~\emph{equals} $W_1(\mu_i,\nu)$. This guarantees that Flowtree recovers the planted nearest neighbor, and this event depends only on $d,s,\epsilon$, and is independent of $N$.
\end{proof}

\section{Additional Experiments}

\subsection{Additional Sinkhorn and ACT Experiments}

\begin{figure*}[h]
\vskip 0.2in
\begin{center}
\caption{Additional Sinkhorn and ACT experiments}
\medskip
\begin{subfigure}{0.5\textwidth}
\includegraphics[width=1.0\textwidth]{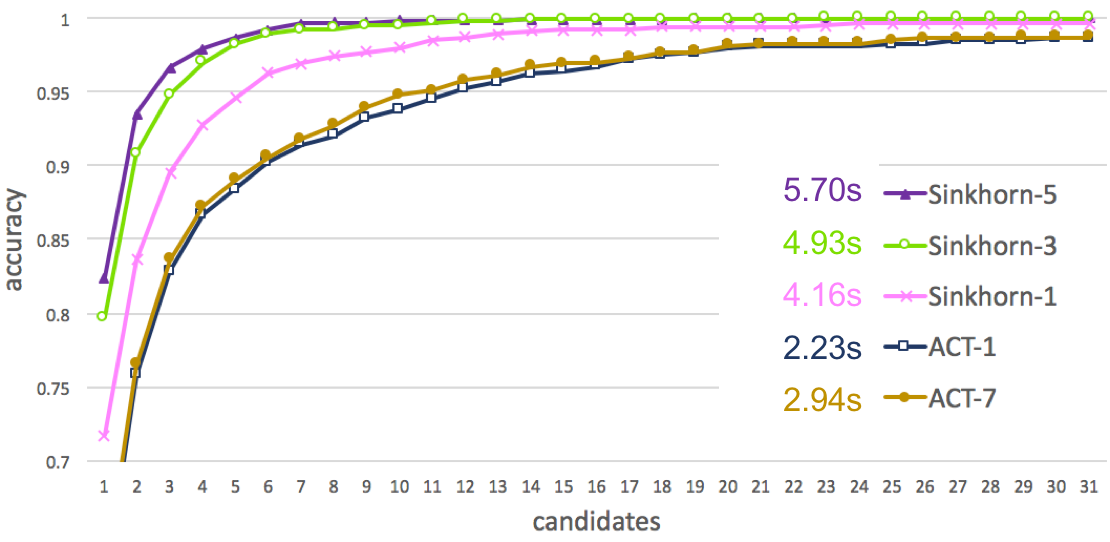}
\caption{20news dataset}
\end{subfigure}%
\begin{subfigure}{0.5\textwidth}
\includegraphics[width=1.0\textwidth]{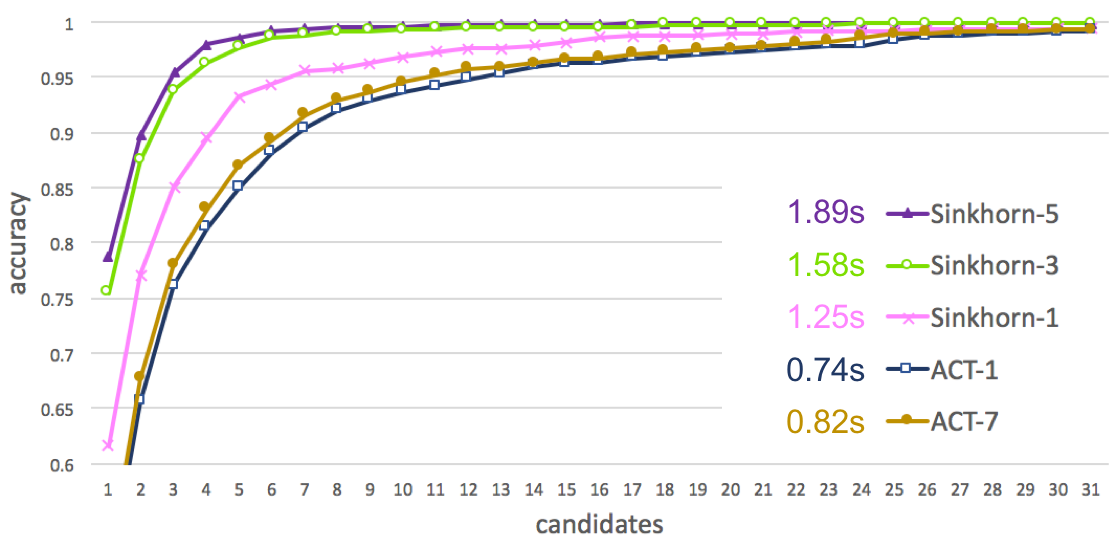}
\caption{Amazon dataset}
\end{subfigure}\\
\begin{subfigure}{0.5\textwidth}
\includegraphics[width=1.0\textwidth]{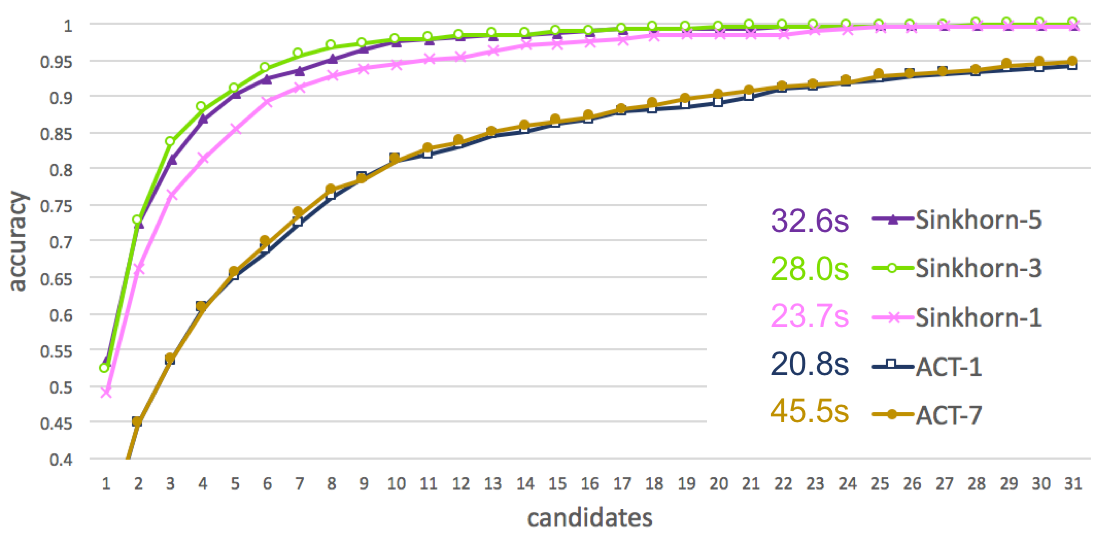}
\caption{MNIST dataset}
\end{subfigure}%
\begin{subfigure}{0.5\textwidth}
\includegraphics[width=1.0\textwidth]{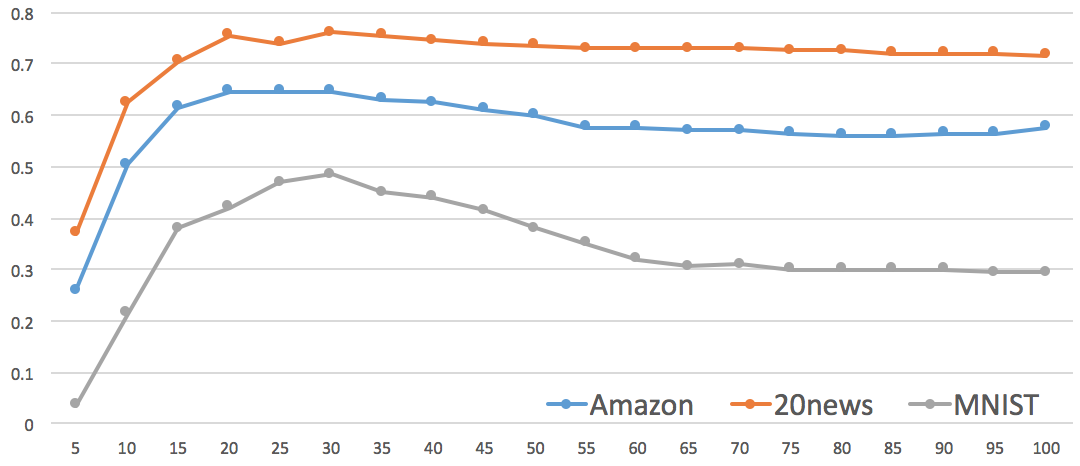}
\caption{$1$-NN accuracy of Sinkhorn-1 with varying regularization}
\end{subfigure}
\label{fig:sink135}
\end{center}
\vskip -0.2in
\end{figure*}

\paragraph{Number of iterations.}
Both ACT and Sinkhorn are iterative algorithms, and the number of iterations is a parameter to set. 
Our main experiments use ACT with 1 iteration and Sinkhorn with 1 or 3 iterations. 
The next experiments motivate these choices. 
Figures~\ref{fig:sink135}(a)--(c)  depict the accuracy and running time of ACT-1, ACT-7, Sinkhorn-1, Sinkhorn-3 and Sinkhorn-5 on each of our datasets.\footnote{ACT-1 and ACT-7 are the settings reported in~\cite{atasu2019linear}.} It can be seen that for both algorithms, increasing the number of iterations beyond the settings used in Section~\ref{sec:experiments} yields comparable accuracy with a slower running time. Therefore in Section~\ref{sec:experiments} we restrict our evaluation to ACT-1, Sinkhorn-1 and Sinkhorn-3. 
We also remark that in the pipeline experiments, we have evaluated Sinkhorn with up to 9 iterations. In those experiments too, the best results are achieved with either 1 or 3 iterations


\paragraph{Sinkhorn regularization parameter.}
Sinkhorn has a regularization parameter $\lambda$ that needs to be tuned per dataset.
We set $\lambda=\eta\cdot M$, where $M$ is the maximum value in the cost matrix (of the currently evaluated pair of distributions), and tune $\eta$. In all of our three datasets the optimal setting is $\eta=30$, which is the setting we use in Section~\ref{sec:experiments}.
As an example, Figure~\ref{fig:sink135}(d) depicts the $1$-NN accuracy (y-axis) of Sinkhorn-1 per $\eta$ (x-axis).

\subsection{Additional Pipeline Results}
The next tables summarize the running times and parameters settings of all pipelines considered in our experiments (whereas the main text focuses on pipelines that start with Quadtree, since it is superior as a first step to Mean and Overlap). The listed parameters are the number of output candidates of each step in the pipeline.

In the baseline pipelines, parameters are tuned to achieve optimal performance (i.e., minimize the running time while attaining the recall goal on at least 90\% of the queries). The details of the tuning procedure is as follows. For all pipelines we use the same random subset of $300$ queries for tuning. Suppose the pipeline has $\ell$ algorithms. For $i=1,\ldots,\ell$, let $c_i$ the output number of candidates of the $i$th algorithm in the pipeline. Note that $c_\ell$ always equals either $1$ or $5$, according to the recall goal of the pipeline, so we need to set $c_1,\ldots,c_{\ell-1}$. Let $p_1$ be the recall@1 accuracy of the first algorithm in the pipeline. Namely, $p_1$ is the fraction of queries such that the top-ranked $c_1$ candidates by the first algorithm contain the true nearest neighbor. We calculate 10 possible values of $c_1$, corresponding to $p_1\in\{0.9, 0.91, \ldots, 0.99\}$. We optimize the pipeline by a full grid search over those values of $c_1$ and all possible values of $c_2,\ldots,c_{\ell-1}$.

When introducing Flowtree into a pipeline as an intermediate method, we do not re-optimize the parameters, but rather set its output number of candidates to the maximum between $10$ and twice the output number of candidates of the subsequent algorithm in the pipeline. Re-optimizing the parameters could possibly improve results.

\begin{table*}[h]
\vspace{30pt}
\centering
\begin{tabular}{|l|l|l|}
\hline
Pipeline methods            & Candidates  & Time  \\ \hline
Mean, Sinkhorn-1, Exact     & 1476, 11, 1  & 0.543 \\ \hline
Mean, Sinkhorn-3, Exact     & 1476, 5, 1  & 0.598 \\ \hline
Mean, R-WMD, Exact           & 1850, 28, 1 & 0.428 \\ \hline
Mean, ACT-1, Exact          & 1677, 14, 1 & 0.420 \\ \hline
Overlap, Sinkhorn-1, Exact  & 391, 6, 1   & 0.610 \\ \hline
Overlap, Sinkhorn-3, Exact  & 391, 5, 1   & 0.691 \\ \hline
Overlap, R-WMD, Exact        & 576, 14, 1  & 0.367 \\ \hline
Overlap, ACT-1, Exact       & 434, 10, 1  & 0.429 \\ \hline
Quadtree, Sinkhorn-1, Exact & 295, 5, 1   & 0.250 \\ \hline
Quadtree, Sinkhorn-3, Exact & 227, 3, 1   & 0.248 \\ \hline
Quadtree, R-WMD, Exact       & 424, 12, 1  & 0.221 \\ \hline
Quadtree, ACT-1, Exact      & 424, 8, 1   & 0.236 \\ \hline
\end{tabular}
\caption{\label{tab:1}Recall@1, no Flowtree.}
\end{table*}

\begin{table*}[h]
\centering
\begin{tabular}{|l|l|l|}
\hline
Pipeline methods                      & Candidates      & Time  \\ \hline
Mean, {\bf Flowtree}, Sinkhorn-1, Exact     & 1850, 10, 5, 1  & 0.089 \\ \hline
Mean, {\bf Flowtree}, Sinkhorn-3, Exact     & 1677, 10, 4, 1  & 0.077 \\ \hline
Mean, {\bf Flowtree}, R-WMD, Exact           & 2128, 48, 24, 1 & 0.242 \\ \hline
Mean, {\bf Flowtree}, ACT-1, Exact          & 2128, 20, 10, 1 & 0.138 \\ \hline
Overlap, {\bf Flowtree}, Sinkhorn-1, Exact  & 489, 10, 5, 1   & 0.087 \\ \hline
Overlap, {\bf Flowtree}, Sinkhorn-3, Exact  & 576, 10, 3, 1    & 0.076 \\ \hline
Overlap, {\bf Flowtree}, R-WMD, Exact        & 576, 28, 14, 1  & 0.173 \\ \hline
Overlap, {\bf Flowtree}, ACT-1, Exact       & 576, 16, 8, 1  & 0.119 \\ \hline
Quadtree, {\bf Flowtree}, Sinkhorn-1, Exact & 424, 10, 5, 1   & 0.074 \\ \hline
Quadtree, {\bf Flowtree}, Sinkhorn-3, Exact & 424, 10, 3, 1   & 0.059 \\ \hline
Quadtree, {\bf Flowtree}, R-WMD, Exact       & 424, 22, 11, 1  & 0.129 \\ \hline
Quadtree, {\bf Flowtree}, ACT-1, Exact      & 424, 16, 8, 1   & 0.104 \\ \hline
Mean, {\bf Flowtree}, Exact                 & 1850, 9, 1      & 0.105 \\ \hline
Overlap, {\bf Flowtree}, Exact              & 489, 9, 1       & 0.100 \\ \hline
Quadtree, {\bf Flowtree}, Exact             & 424, 9, 1       & 0.092 \\ \hline
\end{tabular}
\caption{\label{tab:2}Recall@1, with Flowtree.}
\end{table*}

\begin{table*}[]
\centering
\begin{tabular}{|l|l|l|}
\hline
Pipeline methods       & Candidates & Time  \\ \hline
Mean, Sinkhorn-1       & 1476, 5    & 0.464 \\ \hline
Mean, Sinkhorn-3       & 1476, 5    & 0.549 \\ \hline
Mean, R-WMD, Exact & 1850, 28, 5 & 0.426 \\ \hline
Mean, ACT-1, Exact & 1677, 14, 5 & 0.423 \\ \hline
Overlap, Sinkhorn-1    & 391, 5     & 0.560 \\ \hline
Overlap, Sinkhorn-3    & 391, 5     & 0.650 \\ \hline
Overlap, R-WMD, Exact    & 576, 14, 5     & 0.368 \\ \hline
Overlap, ACT-1, Exact    & 434, 10, 5     & 0.428 \\ \hline
Quadtree, Sinkhorn-1   & 295, 5     & 0.222 \\ \hline
Quadtree, Sinkhorn-3   & 227, 5     & 0.200 \\ \hline
Quadtree, R-WMD, Exact & 424, 11, 5 & 0.216 \\ \hline
Quadtree, ACT-1, Exact & 424, 7, 5  & 0.222 \\ \hline
\end{tabular}
\caption{\label{tab:3}Recall@5, no Flowtree.}
\end{table*}

\begin{table*}[]
\centering
\begin{tabular}{|l|l|l|}
\hline
Pipeline methods                 & Candidates     & Time  \\ \hline
Mean, {\bf Flowtree}, Sinkhorn-1       & 1850, 10, 5    & 0.046 \\ \hline
Mean, {\bf Flowtree}, Sinkhorn-3       & 1476, 10, 5    & 0.043 \\ \hline
Mean, {\bf Flowtree}, R-WMD, Exact       & 2128, 48, 24, 5    & 0.237 \\ \hline
Mean, {\bf Flowtree}, ACT-1       & 2128, 10, 5    & 0.048 \\ \hline
Overlap, {\bf Flowtree}, Sinkhorn-1    & 391, 10, 5     & 0.042 \\ \hline
Overlap, {\bf Flowtree}, Sinkhorn-3    & 391, 10, 5     & 0.044 \\ \hline
Overlap, {\bf Flowtree}, R-WMD, Exact    & 576, 28, 14, 5    & 0.173 \\ \hline
Overlap, {\bf Flowtree}, ACT-1       & 576, 10, 5    & 0.046 \\ \hline
Quadtree, {\bf Flowtree}, Sinkhorn-1   & 424, 10, 5     & 0.033 \\ \hline
Quadtree, {\bf Flowtree}, Sinkhorn-3   & 424, 10, 5     & 0.034 \\ \hline
Quadtree, {\bf Flowtree}, ACT-1        & 424, 10, 5     & 0.029 \\ \hline
Mean, {\bf Flowtree}                   & 2128, 5        & 0.043 \\ \hline
Overlap, {\bf Flowtree}                & 576, 5         & 0.039 \\ \hline
Quadtree, {\bf Flowtree}               & 645, 5         & 0.027 \\ \hline
Quadtree, {\bf Flowtree}, R-WMD, Exact & 424, 22, 11, 5 & 0.131 \\ \hline
Quadtree, {\bf Flowtree}, ACT-1, Exact & 424, 16, 8, 5  & 0.103 \\ \hline
\end{tabular}
\caption{\label{tab:4}Recall@5, with Flowtree.}
\end{table*}

\end{document}